\documentclass[12pt]{article}
\usepackage{graphicx}

\newcommand\pubdate{\today}

\newcommand\pubnumber{LHCb-PROC-2011-044}

\def\Title#1{\begin{center} {\Large #1 } \end{center}}
\def\Author#1{\begin{center}{ \sc #1} \end{center}}
\def\Address#1{\begin{center}{ \it #1} \end{center}}

\newcommand\pubblock{\rightline{\begin{tabular}{l} \pubnumber\\
         \pubdate  \end{tabular}}}
\newenvironment{Abstract}{\begin{center}{\bf Abstract}\end{center} \bigskip \begin{quotation}  }{\end{quotation}}
\newenvironment{Presented}{\begin{quotation} \begin{center} 
             PRESENTED AT\end{center}\bigskip 
      \begin{center}\begin{large}}{\end{large}\end{center} \end{quotation}}

\let\bar=\overbar

\def\Dslash{\not{\hbox{\kern-4pt $D$}}}
\def\dslash{\not{\hbox{\kern-2pt $\del$}}}

\def\msb{{\bar{\ssstyle M \kern -1pt S}}}

\begin{document}
\begin{titlepage}
\pubblock

\vfill


\Title{LHCb time-dependent results}
\vfill
\Author{Marta Calvi}  
\Address{University of Milano-Bicocca, Milano, I-20126 Italy}
\center{On behalf of the LHCb Collaboration}
\vfill


\begin{Abstract}
This review reports preliminary results of time-dependent measurements of decays of 
${\ensuremath{\rm {B^0}}}$ mesons and ${\ensuremath{\rm {B^0_s}}}$ mesons 
coming from the analysis of about 36 pb$^{-1}$ of 
data collected by the LHCb experiment during the 2010 run of the Large Hadron Collider
at {\rm $\sqrt{s}=7$ TeV}.
\end{Abstract}

\vfill

\begin{Presented}
The Ninth International Conference on\\
Flavor Physics and CP Violation\\
(FPCP 2011)\\
Maale Hachamisha, Israel,  May 23--27, 2011
\end{Presented}
\vfill

\end{titlepage}
\def\thefootnote{\fnsymbol{footnote}}
\setcounter{footnote}{0}
%


\newcommand{\comment}[1]{\par\noindent {\em\small [#1]}}
\renewcommand{\comment}[1]{}


\newcommand{\mysubsection}[1]{\subsection[#1]{\boldmath #1}}

\newcommand{\unit}[1]{\ensuremath{\rm\,#1}}
\newcommand{\keV}{\unit{keV}}
\newcommand{\keVc}{\unit{keV/{\it c}}}
\newcommand{\keVcc}{\unit{keV/{\it c}^2}}
\newcommand{\MeV}{\unit{MeV}}
\newcommand{\MeVc}{\unit{MeV\!/\!{\it c}}}
\newcommand{\MeVcc}{\unit{MeV\!/\!{\it c}^2}}
\newcommand{\MeVcca}{$\rm MeV\!/\!{\it c}^2$}
\newcommand{\GeV}{\unit{GeV}}
\newcommand{\GeVc}{\unit{GeV\!/\!{\it c}}}
\newcommand{\GeVcc}{\unit{GeV\!/\!{\it c}^2}}
\newcommand{\TeV}{\unit{TeV}}
\newcommand{\invfb}{\unit{fb^{-1}}}
\newcommand{\invpb}{\unit{pb^{-1}}}
\newcommand{\cm}{\unit{cm}}
\newcommand{\mm}{\unit{mm}}
\newcommand{\mb}{\unit{mb}}
\newcommand{\ps}{\unit{ps}}
\newcommand{\invps}{\unit{ps^{-1}}}
\newcommand{\invpsa}{$\rm ps^{-1}$}
\newcommand{\fs}{\unit{fs}}
\newcommand{\microns}{\unit{\mu m}}
\newcommand{\mrad}{\unit{mrad}}
\newcommand{\rad}{\unit{rad}}
\newcommand{\Hz}{\unit{Hz}}
\newcommand{\kHz}{\unit{kHz}}
\newcommand{\MHz}{\unit{MHz}}



\newcommand{\CP}{\ensuremath{\rm CP}}
\newcommand{\bra}[1]{\ensuremath{\langle #1|}}
\newcommand{\ket}[1]{\ensuremath{|#1\rangle}}
\newcommand{\braket}[2]{\ensuremath{\langle #1|#2\rangle}}

\newcommand{\IP}{\ensuremath{\rm IP}}
\newcommand{\signif}[1]{\ensuremath{\rm #1/\sigma_{#1}}}
\newcommand{\signifm}[1]{\ensuremath{#1/\sigma_{#1}}}
\newcommand{\pT}{\ensuremath{p_{\rm T}}}
\newcommand{\pTmin}{\ensuremath{\pT^{\rm min}}}
\newcommand{\ET}{\ensuremath{E_{\rm T}}}
\newcommand{\DLL}[2]{\ensuremath{\rm \Delta  \ln{\cal L}_{\particle{#1}\particle{#2}}}}
\newcommand{\BRvis}[1]{\ensuremath{{\cal B}_{\rm vis} \rm (#1)}}

\newcommand{\dm}{\ensuremath{\Delta m}}
\newcommand{\dms}{\ensuremath{\Delta m_{\rm s}}}
\newcommand{\dmd}{\ensuremath{\Delta m_{\rm d}}}
\newcommand{\DG}{\ensuremath{\Delta\Gamma}}
\newcommand{\DGs}{\ensuremath{\Delta\Gamma_{\rm s}}}
\newcommand{\DGd}{\ensuremath{\Delta\Gamma_{\rm d}}}
\newcommand{\Gs}{\ensuremath{\Gamma_{\rm s}}}
\newcommand{\Gd}{\ensuremath{\Gamma_{\rm d}}}

\newcommand{\MBq}{\ensuremath{M_{\Bq}}}
\newcommand{\DGq}{\ensuremath{\Delta\Gamma_{\rm q}}}
\newcommand{\Gq}{\ensuremath{\Gamma_{\rm q}}}
\newcommand{\dmq}{\ensuremath{\Delta m_{\rm q}}}
\newcommand{\GL}{\ensuremath{\Gamma_{\rm L}}}
\newcommand{\GH}{\ensuremath{\Gamma_{\rm H}}}

\newcommand{\DGsGs}{\ensuremath{\Delta\Gamma_{\rm s}/\Gamma_{\rm s}}}
\newcommand{\Dm}{\mbox{$\Delta m $}}
\newcommand{\ACP}{\ensuremath{{\cal A}^{\rm CP}}}
\newcommand{\Adir}{\ensuremath{{\cal A}^{\rm dir}}}
\newcommand{\Amix}{\ensuremath{{\cal A}^{\rm mix}}}
\newcommand{\ADelta}{\ensuremath{{\cal A}^\Delta}}
\newcommand{\phid}{\ensuremath{\phi_{\rm d}}}
\newcommand{\sinphid}{\ensuremath{\sin\!\phid}}
\newcommand{\phis}{\ensuremath{\phi_{\rm s}}}
\newcommand{\betas}{\ensuremath{\beta_{\rm s}}}
\newcommand{\sbetas}{\ensuremath{\sigma(\beta_{\rm s})}}
\newcommand{\stbetas}{\ensuremath{\sigma(2\beta_{\rm s})}}
\newcommand{\stphis}{\ensuremath{\sigma(\phi_{\rm s})}}
\newcommand{\sinphis}{\ensuremath{\sin\!\phis}}

\newcommand{\vOmega}{\ensuremath{\vec{\Omega}}}
\newcommand{\vkappa}{\ensuremath{\vec{\kappa}}}
\newcommand{\vlambda}{\ensuremath{\vec{\lambda}}}

\newcommand{\Af}{\ensuremath{A_{f}}}
\newcommand{\Afbar}{\ensuremath{\overline{A}_{\overline{f}}}}
\newcommand{\fbar}{\ensuremath{\BAR{ f}}}

\newcommand{\calH}{\ensuremath{{\cal H}}}
\newcommand{\tauBs}{\ensuremath{\tau_\Bs}}
\newcommand{\tauL}{\ensuremath{\tau_{\rm L}}}
\newcommand{\tauH}{\ensuremath{\tau_{\rm H}}}

\newcommand{\effs}{{\ensuremath{\varepsilon^{\rm s}_\theta }}}
\newcommand{\effb}{{\ensuremath{\varepsilon^{\rm b}_\theta}}}

\newcommand{\Vud}{\ensuremath{V_{\rm ud}}}
\newcommand{\Vus}{\ensuremath{V_{\rm us}}}
\newcommand{\Vub}{\ensuremath{V_{\rm ub}}}
\newcommand{\Vcd}{\ensuremath{V_{\rm cd}}}
\newcommand{\Vcs}{\ensuremath{V_{\rm cs}}}
\newcommand{\Vcb}{\ensuremath{V_{\rm cb}}}
\newcommand{\Vtd}{\ensuremath{V_{\rm td}}}
\newcommand{\Vts}{\ensuremath{V_{\rm ts}}}
\newcommand{\Vtb}{\ensuremath{V_{\rm tb}}}
\newcommand{\VCKM}{\ensuremath{V_{\rm CKM}}}

\newcommand{\Vjr}{\ensuremath{V_{ jr}}}
\newcommand{\Vjb}{\ensuremath{V_{j {\rm b}}}}

\newcommand{\edet}{{\ensuremath{\varepsilon_{\rm det}}}}
\newcommand{\erec}{{\ensuremath{\varepsilon_{\rm rec/det}}}}
\newcommand{\esel}{{\ensuremath{\varepsilon_{\rm sel/rec}}}}
\newcommand{\etrg}{{\ensuremath{\varepsilon_{\rm trg/sel}}}}
\newcommand{\etot}{{\ensuremath{\varepsilon_{\rm tot}}}}

\newcommand{\mistag}{\ensuremath{\omega}}
\newcommand{\wcomb}{\ensuremath{\omega^{\rm comb}}}
\newcommand{\etag}{{\ensuremath{\varepsilon_{\rm tag}}}}
\newcommand{\etagcomb}{{\ensuremath{\varepsilon_{\rm tag}^{\rm comb}}}}
\newcommand{\effeff}{\ensuremath{\varepsilon_{\rm eff}}}
\newcommand{\effeffcomb}{\ensuremath{\varepsilon_{\rm eff}^{\rm comb}}}
\newcommand{\efftag}{{\ensuremath{\etag(1-2\omega)^2}}}
\newcommand{\effD}{{\ensuremath{\etag D^2}}}

\newcommand{\etagprompt}{{\ensuremath{\varepsilon_{\rm tag}^{\rm Pr}}}} 
\newcommand{\etagLL}{{\ensuremath{\varepsilon_{\rm tag}^{\rm LL}}}}

\newcommand{\tage}{\particle{e}}
\newcommand{\tagmu}{\particle{\mu}}
\newcommand{\tagKopp}{\particle{K_{opp}}}
\newcommand{\tagKsame}{\particle{K_{same}}}
\newcommand{\tagPsame}{\particle{\pi_{same}}}
\newcommand{\tagQvtx}{\ensuremath{Q_{\rm vtx}}}
\newcommand {\mch}      {\multicolumn {4} {|c|}}

\newcommand{\effLO}{\ensuremath{\varepsilon_{\rm L0/sel}}}
\newcommand{\effpresel}{\ensuremath{\varepsilon_{\rm presel/gen}}}
\newcommand{\effsel}{\ensuremath{\varepsilon_{\rm sel/presel}}}
\newcommand{\effgen}{\ensuremath{\varepsilon_{\rm gen}}}


\newcommand{\BAR}[1]{\overline{#1}}

\newcommand{\particle}[1]{{\ensuremath{\rm #1}}}

\newcommand{\pp}{\particle{pp}}
\newcommand{\ppbar}{\particle{p\BAR{p}}}

\newcommand{\cc}{\particle{c\BAR{c}}}
\renewcommand{\b}{\particle{b}}
\newcommand{\bbar}{\particle{\BAR{b}}}
\newcommand{\bb}{\particle{b\BAR{b}}}
\newcommand{\dd}{\particle{d\BAR{d}}}

\newcommand{\B}{\particle{B}}
\newcommand{\Bd}{\particle{B^0}}
\newcommand{\Bs}{\particle{B^0_s}}
\newcommand{\Bds}{\particle{B^0_{(s)}}}
\newcommand{\Bu}{\particle{B^+}}
\newcommand{\Bc}{\particle{B^+_c}}
\newcommand{\Lb}{\particle{\Lambda_b}}

\newcommand{\Bbar}{\particle{\BAR{B}}}
\newcommand{\Bdbar}{\particle{\BAR{B}{^0}}}
\newcommand{\Bsbar}{\particle{\BAR{B}{^0_s}}}
\newcommand{\Bdsbar}{\particle{\BAR{B}{^0}_{(s)}}}
\newcommand{\Bubar}{\particle{B^-}}
\newcommand{\Bcbar}{\particle{B^-_c}}
\newcommand{\Lbbar}{\particle{\BAR{\Lambda}_b}}

\newcommand{\Bqq}{\particle{B_q}}
\newcommand{\Bqqbar}{\particle{\BAR{B_q}}}

\newcommand{\BL}{\particle{B_L}}
\newcommand{\BH}{\particle{B_H}}
\newcommand{\BLH}{\particle{B_{L,H}}}

\newcommand{\Ds}{\particle{D_s}}
\newcommand{\Dsm}{\particle{D_s^-}}
\newcommand{\KKpim}{\particle{K^+K^-\pi^-}}
\newcommand{\KK}{\particle{K^+K^-}}
\newcommand{\Dsp}{\particle{D_s^+}}
\newcommand{\Dsmp}{\particle{D_s^{\mp}}}

\newcommand{\Bq}{\particle{B_q}}
\newcommand{\Bqbar}{\particle{\BAR{B}_{q}}}

\newcommand{\Dz}{\particle{D^0}}
\newcommand{\Dzbar}{\particle{\BAR{D}{^0}}}
\newcommand{\DzCP}{\particle{D^0_{CP}}}
\newcommand{\Dstar}{\particle{D^{*-}}}
\newcommand{\Dstarz}{\particle{D^{0*}}}
\newcommand{\DstInc}{\particle{D^{(*)-}}}

\newcommand{\Jpsi}{\particle{J\!/\!\psi}}
\newcommand{\Jmm}{\particle{\Jpsi(\mu\mu)}}
\newcommand{\Jee}{\particle{\Jpsi(ee)}}
\newcommand{\JpsiX}{\particle{\Jpsi X}}

\newcommand{\KS}{\particle{K^0_S}}  
\newcommand{\Kst}{\particle{K^{*0}}}  
\newcommand{\Kstar}{\particle{K^{*}}}  
\newcommand{\Kstbar}{\particle{\BAR{K}^{*0}}}  

\newcommand{\Kplus}{\particle{K^+}}
\newcommand{\Kminus}{\particle{K^-}}
\newcommand{\Kpm}{\particle{K^\pm}}

\newcommand{\pip}{\particle{\pi^+}}
\newcommand{\pim}{\particle{\pi^-}}
\newcommand{\piz}{\particle{\pi^0}}

\newcommand{\Jphi}{\particle{J\!/\!\psi\phi}}
\newcommand{\BsBs}{\Bs--\Bsbar}
\newcommand{\BB}{\B--\Bbar}


\newcommand{\decay}[2]{\particle{#1\!\to #2}}

\newcommand{\bccsbar}{\decay{\BAR{b}}{\BAR{c}c\BAR{s}}}
\newcommand{\bsssbar}{\decay{\BAR{b}}{\BAR{s}s\BAR{s}}}

\newcommand{\KSpipi}{\decay{K^0_S}{\pi^+\pi^-}}  
\newcommand{\Jpsiee}{\decay{\Jpsi}{e^+e^-}}  
\newcommand{\Jpsimm}{\decay{\Jpsi}{\mu\mu}}  
\newcommand{\Jpsill}{\decay{\Jpsi}{\ell^+\ell^-}}  

\newcommand{\Kpi}{\particle{K^+\pi^-}}
\newcommand{\pippim}{\particle{\pi^+\pi^-}}

\newcommand{\DsKKpi}{\decay{\Dsm}{K^+K^-\pi^-}}  
\newcommand{\phiKK}{\decay{\phi}{K^+K^-}}
\newcommand{\KstKpi}{\decay{\Kst}{\Kpi}}

\newcommand{\Bdpipi}{\decay{\Bd}{\pi^+\pi^-}}            
\newcommand{\BdKpi}{\decay{\Bd}{K^+\pi^-}}               
\newcommand{\BspiK}{\decay{\Bs}{\pi^+K^-}}               
\newcommand{\BsKK}{\decay{\Bs}{K^+K^-}}                  
\newcommand{\Bhh}{\decay{\Bds}{h^+h^-}}                  
\newcommand{\BsDspi}{\decay{\Bs}{\Dsm\pi^+}}             
\newcommand{\BsDstpi}{\decay{\Bs}{\Dsm \pi^+ \pi^- \pi^+}}             
\newcommand{\BsDsK}{\decay{\Bs}{\Dsmp K^{\pm}}}          
\newcommand{\BsDsmKp}{\decay{\Bs}{\Dsm K^+}}             
\newcommand{\BsDspKm}{\decay{\Bs}{\Dsp K^-}}             
\newcommand{\BsDsh}{\decay{\Bs}{\Dsm h^+}}               
\newcommand{\BdJmmKS}{\decay{\Bd}{\Jmm\KS(\pi\pi)}}              
\newcommand{\BdJeeKS}{\decay{\Bd}{\Jee\KS}}              
\newcommand{\BdJKS}{\decay{\Bd}{\Jpsi\KS}}               
\newcommand{\JKS}{\Jpsi\KS}
\newcommand{\BdbarJKS}{\decay{\Bdbar}{\Jpsi\KS}}

\newcommand{\BdJmmKst}{\decay{\Bd}{\Jmm\Kst(\particle{K}\pi)}}            
\newcommand{\BdJeeKst}{\decay{\Bd}{\Jee\Kst}}            
\newcommand{\BdJKst}{\decay{\Bd}{\Jpsi\Kst}}             
\newcommand{\BuJmmK}{\decay{\Bu}{\Jmm K^+}}              
\newcommand{\BuJeeK}{\decay{\Bu}{\Jee K^+}}              
\newcommand{\BuJK}{\decay{\Bu}{\Jpsi K^+}}               
\newcommand{\BsDsmu}{\decay{\Bs}{\Dsm\mu^+\nu}}          

\newcommand{\BuJX}{\decay{\Bu}{\Jpsi X}}  
\newcommand{\BdJX}{\decay{\Bd}{\Jpsi X}}  
\newcommand{\BsJX}{\decay{\Bs}{\Jpsi X}}  
\newcommand{\LbJX}{\decay{\Lb}{\Jpsi X}}  
\newcommand{\BuJmmX}{\decay{\Bu}{\Jmm X}}  
\newcommand{\BdJmmX}{\decay{\Bd}{\Jmm X}}  
\newcommand{\BsJmmX}{\decay{\Bs}{\Jmm X}}  
\newcommand{\LbJmmX}{\decay{\Lb}{\Jmm X}}  
\newcommand{\bJX}{\decay{b}{\Jpsi X}}  
\newcommand{\BJX}{\decay{B}{\Jpsi X}}  

\newcommand{\BsJmmphi}{\decay{\Bs}{\Jmm\phi(\particle{KK})}}            
\newcommand{\BsJeephi}{\decay{\Bs}{\Jee\phi}}            
\newcommand{\BsJeephiKK}{\decay{\Bs}{\Jee\phi(\particle{KK})}}            
\newcommand{\BsJphi}{\decay{\Bs}{\Jpsi\phi}}             
\newcommand{\Bsmm}{\decay{\Bs}{\mu^+\mu^-}}              
\newcommand{\BdmmKst}{\decay{\Bd}{\mu^+\mu^-\Kst}}       
\newcommand{\BdeeKst}{\decay{\Bd}{e^+e^-\Kst}}           
\newcommand{\BdllKst}{\decay{\Bd}{\ell^+\ell^-\Kst}}     
\newcommand{\BdphiKS}{\decay{\Bd}{\phi\KS}}              
\newcommand{\Bsphiphi}{\decay{\Bs}{\phi\phi}}            
\newcommand{\BsKstKst}{\decay{\Bs}{\Kst\Kst}}           
\newcommand{\BdDstpi}{\decay{\Bd}{D^{*-}\pi^+}}          
\newcommand{\BdDstpiincl}{\decay{\Bd}{D^{*-}(incl)\pi^+}}
\newcommand{\BdDKst}{\decay{\Bd}{\Dz\Kst}}               
\newcommand{\BdDbarKst}{\decay{\Bd}{\BAR{D}{^0}\Kst}}    
\newcommand{\BdDCPKst}{\decay{\Bd}{\DzCP\Kst}}           
\newcommand{\BdDbarKpiKst}{\decay{\Bd}{\Dzbar(K\pi)\Kst}}
\newcommand{\BdDbarKKKst}{\decay{\Bd}{\Dzbar(KK)\Kst}}   
\newcommand{\BdDbarpipiKst}{\decay{\Bd}{\Dzbar(\pi\pi)\Kst}}   
\newcommand{\BdDCPKKKst}{\decay{\Bd}{\DzCP(KK)\Kst}}   
\newcommand{\BdKstrho}{\decay{\Bd}{K^{*0}\rho/\omega}}   
\newcommand{\Bsetacphi}{\decay{\Bs}{\eta_c\phi}}         
\newcommand{\BsetacKKphi}{\decay{\Bs}{\eta_c(4K)\phi}}   
\newcommand{\BsetacpiKphi}{\decay{\Bs}{\eta_c(2pi2K)\phi}}  
\newcommand{\Bsetacpipiphi}{\decay{\Bs}{\eta_c(4pi)\phi}}
\newcommand{\BsJmmeta}{\decay{\Bs}{\Jmm\eta}}            
\newcommand{\BsJeeeta}{\decay{\Bs}{\Jee\eta}}            
\newcommand{\BsJeta}{\decay{\Bs}{\Jpsi\eta}}             
\newcommand{\BdKstgam}{\decay{\Bd}{K^{*0}\gamma}}        
\newcommand{\BdKstpin}{\decay{\Bd}{K^{*0}\pi^0}}         
\newcommand{\Bsphigam}{\decay{\Bs}{\phi\gamma}}          
\newcommand{\Bsphipin}{\decay{\Bs}{\phi\pi^0}}           
\newcommand{\Bdpipipi}{\decay{\Bd}{\pi^+\pi^-\pi^0}}     
\newcommand{\Bdrhopi}{\decay{\Bd}{\rho\pi}}              

\newcommand{\BcJmmpi}{\decay{\Bc}{\Jmm\pi^+}}            
\newcommand{\BcJpi}{\decay{\Bc}{\Jpsi\pi^+}}             
\newcommand{\Bsmumu}{\decay{\Bs}{\mu^+\mu^-}}            

\newcommand{\BsDsstDsst}{\decay{\Bs}{\Ds^{(*)-}\Ds^{(*)+}}}  
\newcommand{\BsDsDs}{\decay{\Bs}{\Dsm\Dsp}}             
\newcommand{\BdDstmu}{\decay{\Bd}{\Dstar \mu^+\nu_\mu}}   %

\newcommand{\BudsJX}{\decay{\rm B_{\rm u,d,s}}{\Jpsi \rm X}}


\newcommand{\SM}{Standard Model}             
\newcommand{\fsig}{\ensuremath{f_{\rm sig}}}

\newcommand{\beq}{\begin{equation}}
\newcommand{\eeq}{\end{equation}}


\newcommand{\mbs}{m_{\Bs}}   

\newcommand{\delone}{\delta_1}
\newcommand{\deltwo}{\delta_2}
\newcommand{\delperp}{\delta_{\perp}}
\newcommand{\delpar}{\delta_{\|}}
\newcommand{\delzero}{\delta_0}

\def\thetatwo{\theta_2}
\def\apar{A_{\|}(t)}
\def\aperp{A_{\perp}(t)}
\def\azero{A_{0}(t)}
\def\aparO{A_{\|}(0)}
\def\aperpO{A_{\perp}(0)}
\def\azeroO{A_{0}(0)}
\newcommand{\Rt}{R_\perp}
\newcommand{\Rp}{R_\|}
\newcommand{\Ro}{R_0}
\newcommand{\dGtwo}{\frac{\DG}{2}}
\renewcommand{\d}{{\rm d}}

\def\aparb{\bar{A}_{\|}(t)}
\def\aperpb{\bar{A}_{\perp}(t)}
\def\azerob{\bar{A}_{0}(t)}
\def\aparbO{\bar{A}_{\|}(0)}
\def\aperpbO{\bar{A}_{\perp}(0)}
\def\azerobO{\bar{A}_{0}(0)}


\newcommand{\thetatr}{\ensuremath{\theta}}
\newcommand{\phitr}{\ensuremath{\varphi}}
\newcommand{\psitr}{\ensuremath{\psi}}
\newcommand{\thetaone}{\ensuremath{\psi}}

\newcommand{\phijphi}{\ensuremath{\phi_{\rm s}^{\Jphi}}}
\newcommand{\Phiphiphi}{\ensuremath{\Phi_{\Bsphiphi}}}
\newcommand{\myphis}{\ensuremath{\phi_{\rm s}}}

\newcommand{\phiM}{\ensuremath{\Phi_{\rm M}}}
\newcommand{\phiD}{\ensuremath{\Phi_{\rm D}}}
\newcommand{\phiCP}{\ensuremath{\Phi_{\rm CP}}}
\newcommand{\phiMNP}{\ensuremath{\Phi^{\rm NP}_{\rm M}}}
\newcommand{\phiDNP}{\ensuremath{\Phi^{\rm NP}_{\rm D}}}
\newcommand{\phiMG}{\ensuremath{\phi_{\rm s}^{\rm M/\Gamma}}}
\newcommand{\phiMGSM}{\ensuremath{\phi^{\rm M/\Gamma}_{\rm s \, ,SM}}}

\newcommand{\phipen}{\ensuremath{\Phi_{\rm penguin}}}
\newcommand{\phibox}{\ensuremath{\Phi_{\rm box}}}

\newcommand{\phiNP}{\ensuremath{\Phi^{\rm NP}}}
\newcommand{\phiSM}{\ensuremath{\phi_{\rm s}^{\rm SM}}}
\newcommand{\phisDelta}{\ensuremath{\phi^\Delta_{\rm s}}}
\newcommand{\phitot}{\ensuremath{\Phi_{\rm tot}}}

\newcommand{\diffX}{ \frac{\d^{4} \Gamma}{\d t\, \d\Omega} }
\newcommand{\diffXbar}{ \frac{\d^{4} \bar \Gamma}{\d t\, \d\Omega} }

\newcommand{\calL}{\ensuremath{{\cal L}}}
\newcommand{\Ps}{\ensuremath{{\cal P}_{s}}}
\newcommand{\Psbar}{\ensuremath{\bar{{\cal P}_{s}}}}

\section{Introduction}

LHCb is a single arm forward detector at the Large Hadron Collider
which covers the region of pseudorapidity $2<\eta<5$ where the production 
of  $\rm b{\overline{\rm b}}$ pairs from $pp$ collisions at {\rm $\sqrt(s)=7$} TeV
is copious. A cross section 
$\sigma(pp\rightarrow bbX) \sim 290$ $\mu b$ 
has been recently measured~\cite{ref:LHCbCross}. 

A full description of the LHCb detector can be found in~\cite{ref:Detectorpaper}.
It is worthwhile to summarize the main aspects of the performance which are crucial for 
the time-dependent measurements presented here: 
an efficient trigger for leptonic and hadronic decay modes (with efficiency of 
about 94\% and 60\% respectively) an excellent resolution for tracking and vertexing 
( eg. a resolution on the impact parameter with respect to the primary vertex
$\sigma_{IP}^x \sim 15$ $\mu$m ) and good particle identification from the RICH detectors, 
Calorimeters and  Muon Chambers.

During the first run of LHC in 2010, LHCb collected about 
36 pb$^{-1}$ of data. The preliminary results related to time-dependent analyses 
of \Bd\ and \Bs\ meson decays are presented in the following.

\section{Constraint on the \Bs\ mixing phase \myphis}  

For a final state $f$ equal to its CP conjugate ($f=\bar f$), 
a CP violating phase arises from 
the interference between  ${\rm B^0_s}$ decay to $f$ either directly or 
via \BsBs\ oscillation, producing a difference 
in the proper-time distributions of \Bs\ mesons and \Bsbar\ mesons.
The \BsJmmphi\ decay is considered the golden mode for measuring this type of 
CP violation.
Within the Standard Model (SM) this decay is dominated by \bccsbar\ 
quark level transitions, neglecting QCD penguin contribution the phase is:
$\phiSM=-2\betas$, where $\betas=\arg\left(-\Vts\Vtb^*/\Vcs\Vcb^*\right)$.
Global fits to experimental data give a small and precise value
$2\betas=(-0.0363\pm0.0017)$ $\rad$~\cite{ref:CKMfit}.
New particles could contribute to the mixing box diagram modifying the SM 
prediction, adding a new phase~\cite{ref:NPbetas}.
Recent results from experiment at Tevatron give hints of deviations 
of \myphis\ from the SM predicted value~\cite{ref:TevtronPhis}, with an uncertainty 
on $\phi_s$ of about 0.5 $\rad$. 

The precise determination of $\phi_s$ is one of the key goals of the LHCb experiment. 
This measurement has been carefully prepared with a series of intermediate studies
to prove the capability of LHCb to perform a clean signal selection 
for \BJX\ channels, to control the proper-time and the angular acceptance and resolutions 
and to have a correct determination of the flavour of the B mesons at production. 
All these steps will be briefly summarised in the following.

\begin{figure}[htb]
\centering
\includegraphics[width=0.4\textwidth]{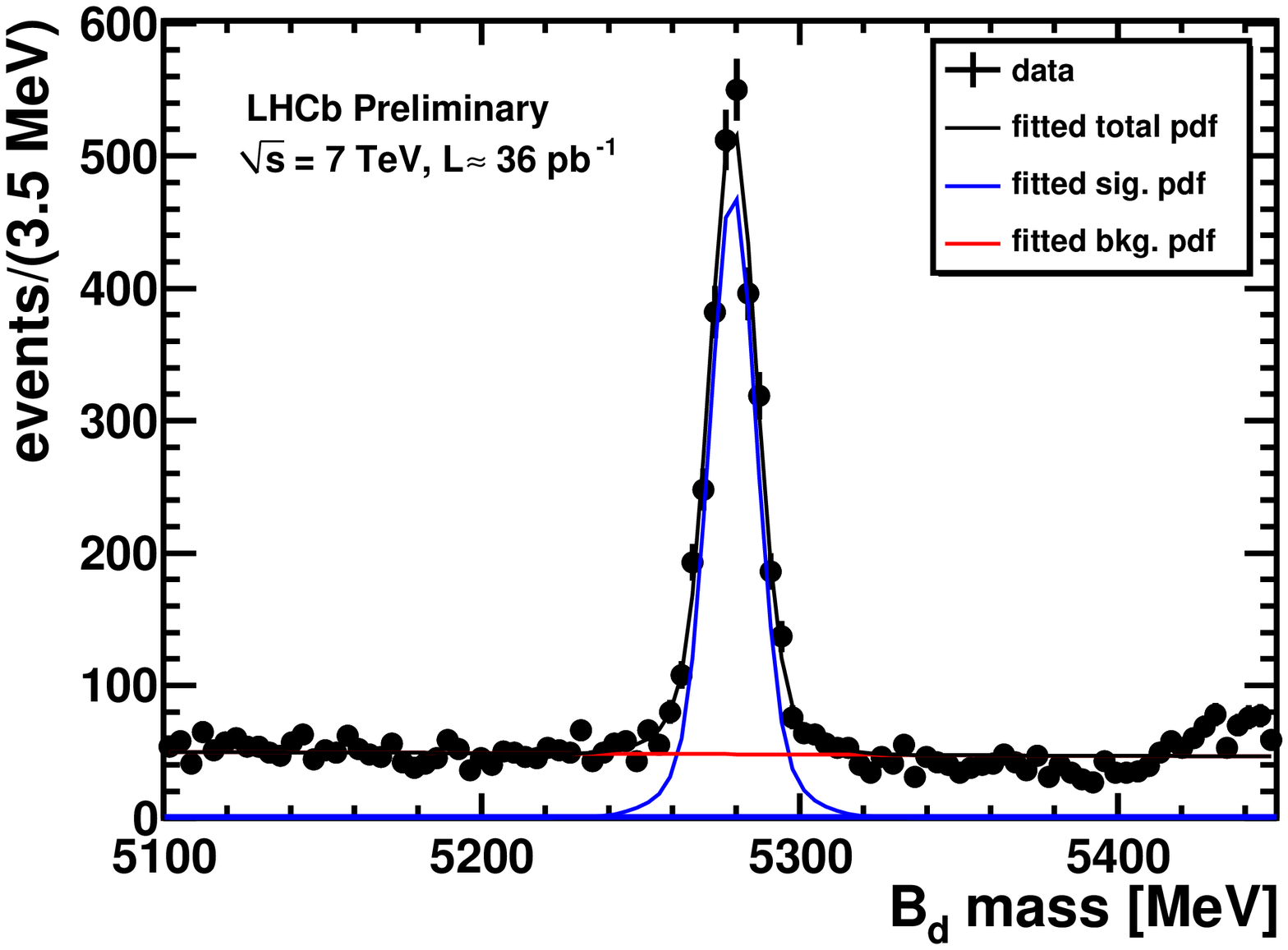}
\includegraphics[width=0.4\textwidth]{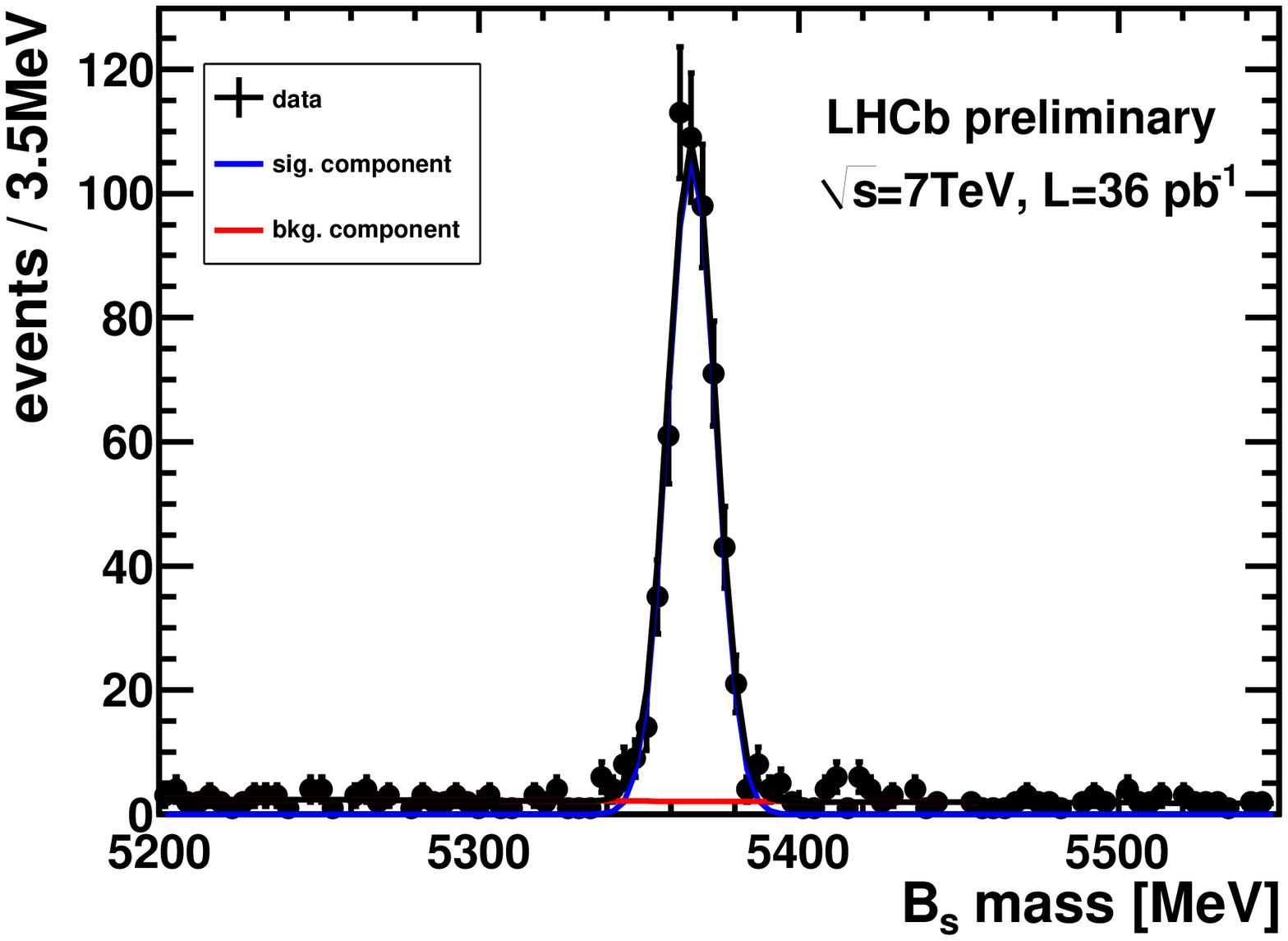}
\caption{Invariant mass distribution of selected \BdJKst\ (left) and  \BsJphi\ (right) 
events.
}
\label{fig:mass}
\end{figure}

A decay-time unbiased di-muon trigger is used to select all
\BJX\ channels is a similar way. Low background remains after the additional 
requirements on the proper-time $t>0.3$ ps, which removes
mainly events with a prompt \Jmm.
As an example the mass distribution of the selected \BdJKst\ and \BsJphi\ 
events are shown in Fig.~\ref{fig:mass}; the mass resolution are 7 and 
8 \MeVcc\,respectively. 

An average proper-time resolution of about 50 ps is determined on data, 
using prompt \Jpsi\ events.
The lifetimes of several b-hadron species have been measured, the 
results shown in Table~\ref{tab:lifetimes}~\cite{ref:note1}
are all compatible with world averages~\cite{ref:PDG} 
The current systematic uncertainties are conservative estimates dominated 
by the uncertainty in the time dependence of the reconstruction efficiency.

\begin{table}[!hbtp]
 \centerline{
   \begin{tabular}{|l|c|r|r|}
      \hline
      Channel   & Lifetime (ps)     & Yield \\\hline\hline
      \BuJK     & 1.689 $\pm$ 0.022 $\pm$ 0.047 & 6741 $\pm$ 85    \\
      \BdJKst   & 1.512 $\pm$ 0.032 $\pm$ 0.042 & 2668 $\pm$ 58    \\
      \BdJKS    & 1.558 $\pm$ 0.056 $\pm$ 0.022 & 838 $\pm$ 31     \\
      \BsJphi   & 1.447 $\pm$ 0.064 $\pm$ 0.056 &570 $\pm$ 24      \\
      $\Lambda_b \to \Jpsi\Lambda$ & 1.353 $\pm$ 0.108 $\pm$ 0.035 & 187 $\pm$ 16    \\
      \hline
      \end{tabular}
  }
  \caption{Signal event yields and lifetimes of different b-hadrons.
The first uncertainty is statistical and the second is systematic.
LHCb preliminary results from 36 \invpb. }
\label{tab:lifetimes}
\end{table}

The decay \BsJphi\ is a pseudo-scalar to vector-vector transition whose
final state is a superposition of three possible states with
relative orbital angular momentum between the vector mesons $\ell=0,1,2$.
In order to disentangle statistically the three components 
an angular analysis of the decay product distributions is required. 
The three angles $\Omega = \{\thetatr, \phitr, \thetaone\}$
defined in the transversity basis are used, as in~\cite{ref:roadmap}. 
The decay can be described by three time-dependent amplitudes, at $t=0$ 
$\azeroO$ and $\aparO$ are CP-even 
while  $\aperpO$ is CP-odd.
The differential decay rate depends on several physics parameters: 
the $\Bs$  decay width \Gs, 
the decay width difference between the two \Bs\ mass eigenstates \DGs, 
the mixing frequency \dms, the CP violating phase \myphis\, 
the relative phases and magnitudes of the three angular transversity amplitudes. 
It should be noted that the dependence on \myphis\ is present in several terms, 
but for small  \myphis\ values, around the SM value,  the main sensitivity comes 
from terms proportional to $\sin\myphis$. Most of these terms are multiplied by 
$\sin(\dms t)$, hence information on  \myphis\ is mainly obtained from observation 
of the amplitude of the fast \Bs\ oscillations in the time distribution.  
These terms have opposite sign between $\Bs$ and $\Bsbar$, 
therefore the analysis benefits significantly from determining the flavour 
of the initial mesons (flavour tagging). 

\begin{figure}[htb]
\centering
\includegraphics[width=0.33\textwidth]{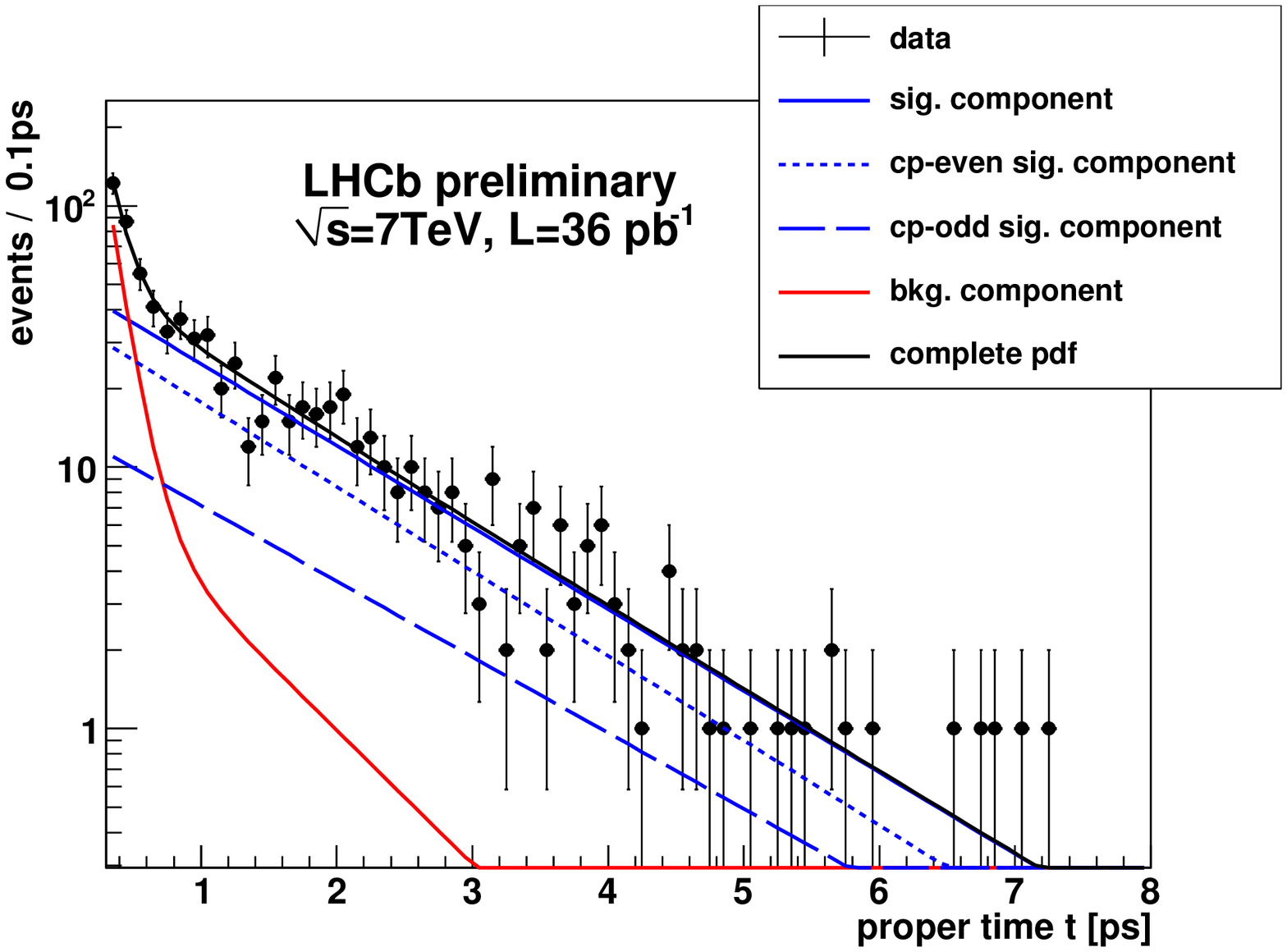}
\includegraphics[width=0.33\textwidth]{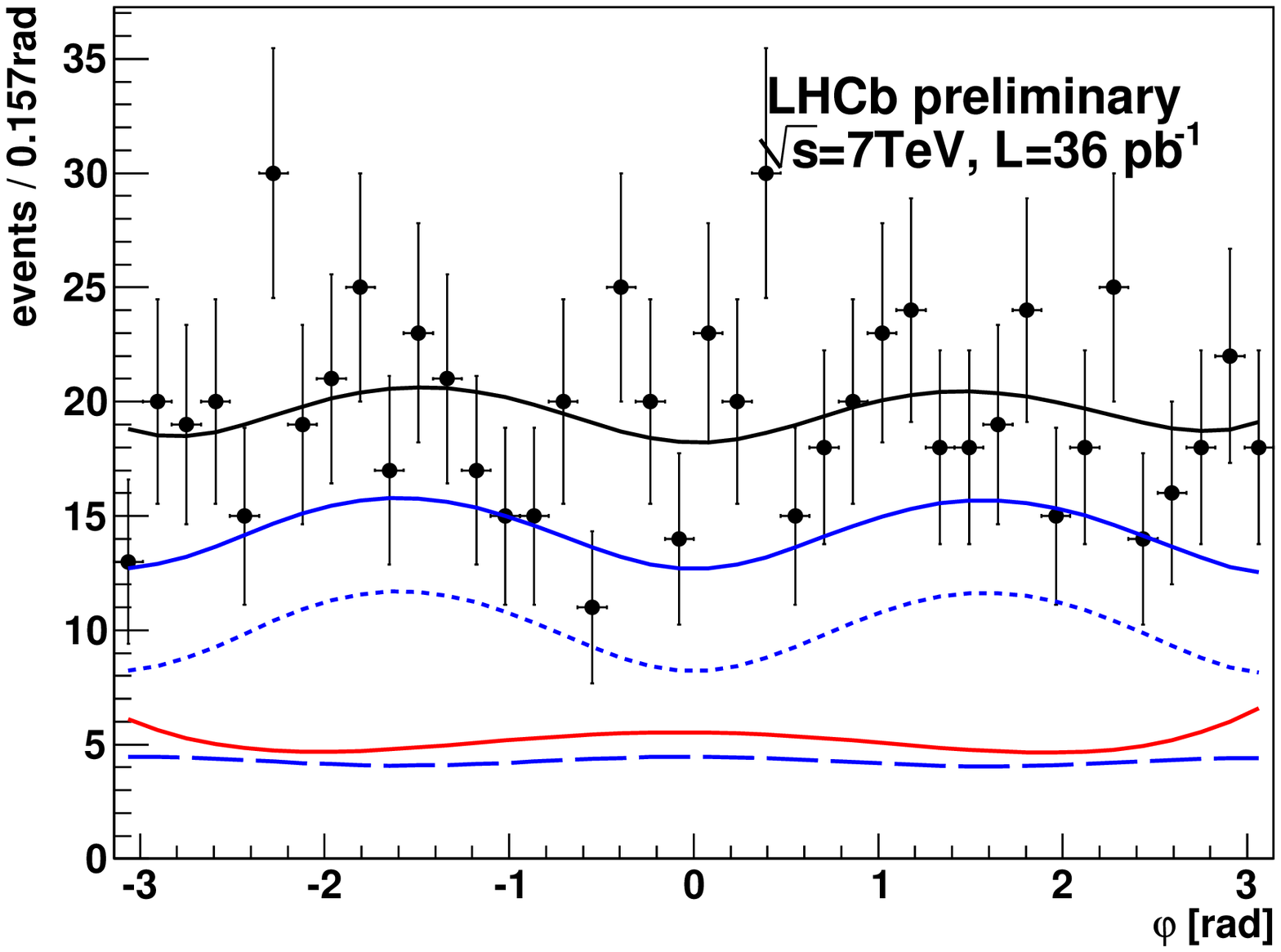}\\
\vskip 0.5cm
\includegraphics[width=0.33\textwidth]{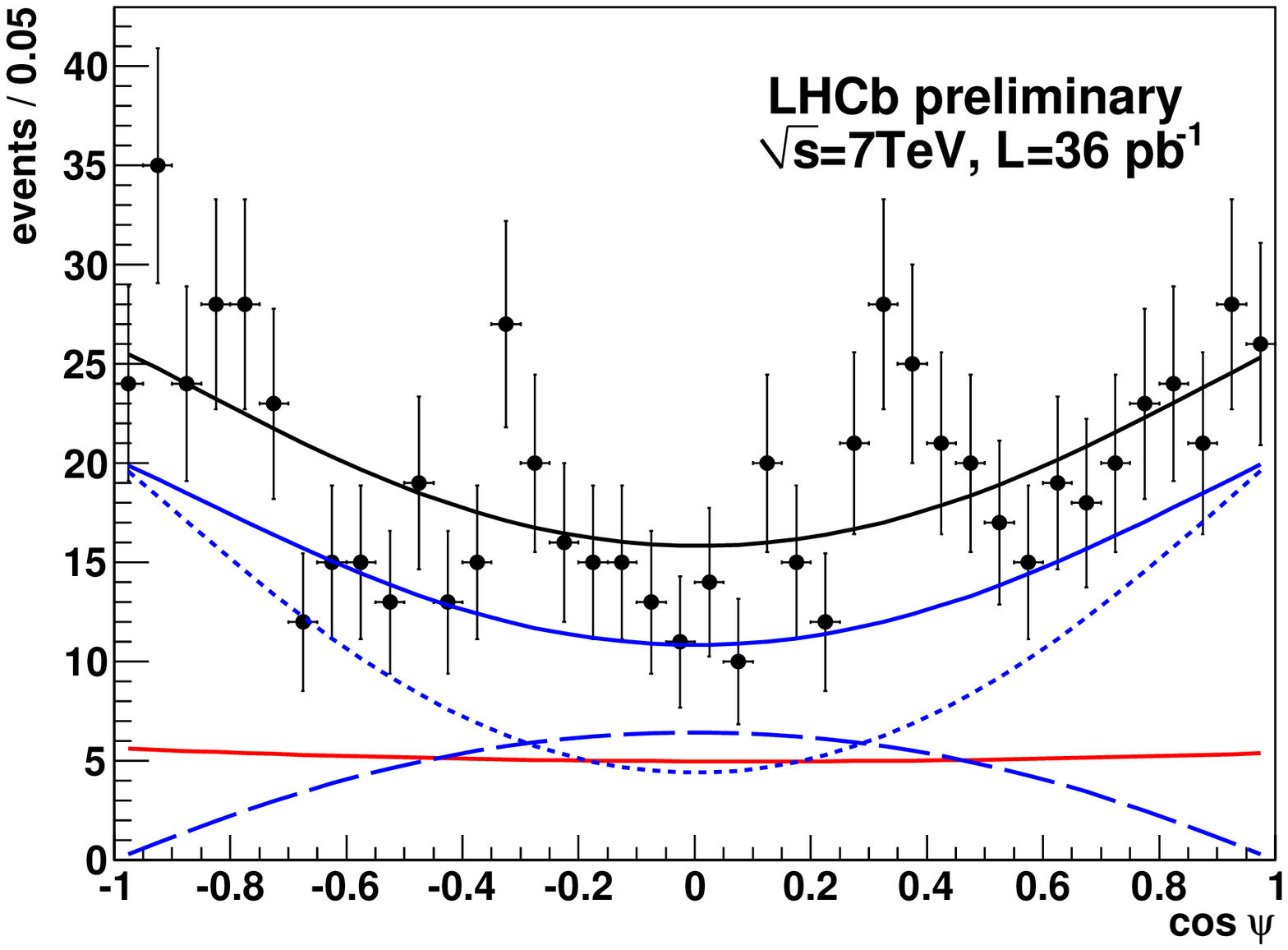}
\includegraphics[width=0.33\textwidth]{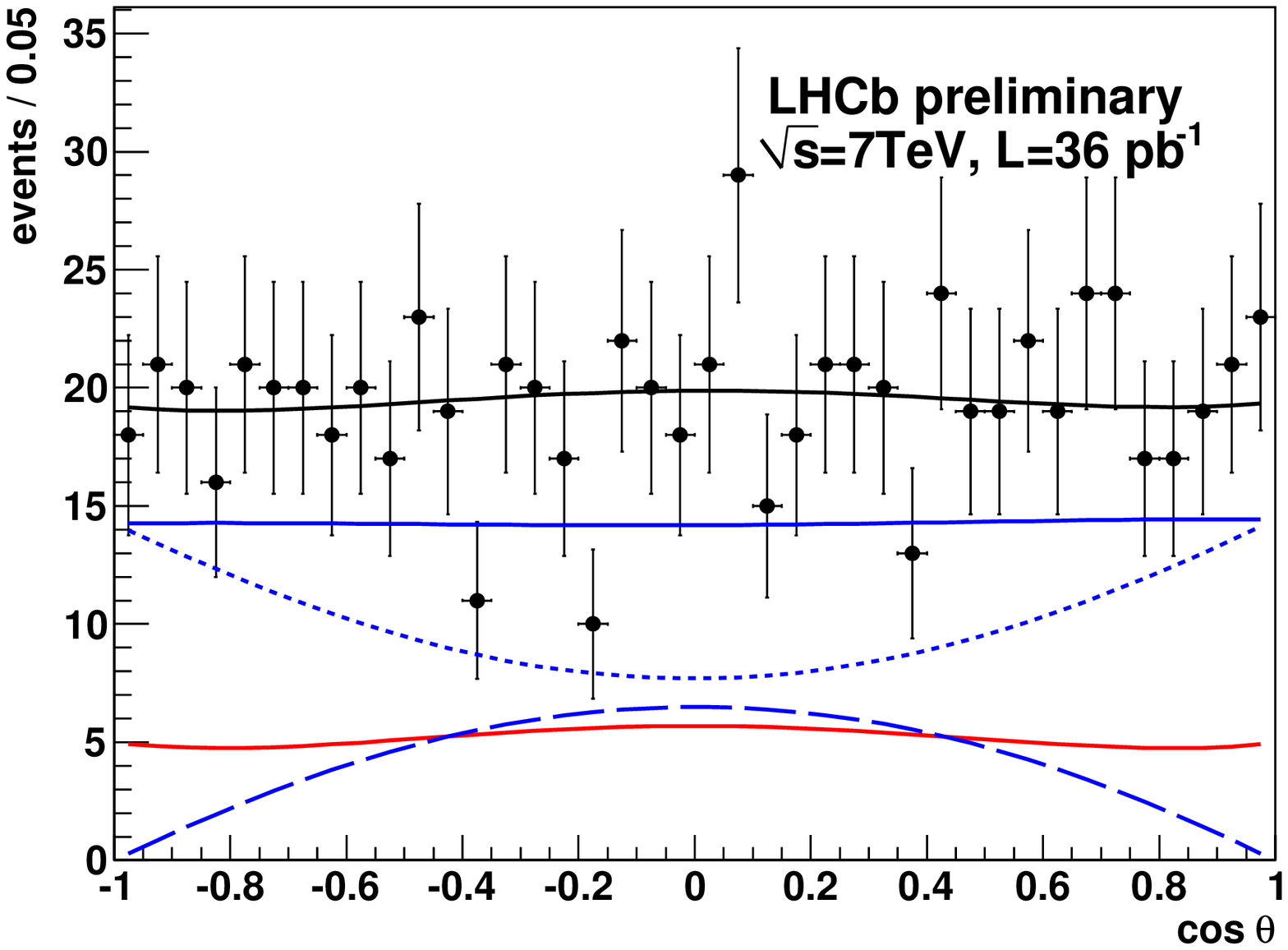}
\caption{Fitted PDF projected on the the proper-time and the transversity angles
compared to the data distributions for the  selected \BsJphi\ candidates
(untagged analysis). Shown are the total PDF, the PDFs for signal, the PDFs for the
CP-even and CP-odd signal components and the total background PDF.}
\label{fig:untagged}
\end{figure}

However, given the additional complication introduced by the tagging procedure, 
and the small number of tagged events, a first analysis is performed 
without using this information ({\it untagged analysis}). 
In this case CP violation is ignored, but from the study of the final state 
angular distributions as a function of proper-time 
the other physics parameters are determined. 
Due to the forward geometry of the LHCb detector the reconstruction efficiency 
for the final state particles is a non-trivial function of the decay angles, 
hence understanding the angular acceptance is an important ingredient of 
this measurement.
The angular acceptance has been studied with Monte Carlo simulated events,
the maximum deviation with respect to the generated angular distributions
results are within about $\pm$5\% .
Results of the likelihood fit to the mass, proper-time and angular distributions 
are shown in Table~\ref{tab:untagged} and 
Fig.~\ref{fig:untagged}~\cite{ref:note2}.

\begin{table}[!hbtp]

\begin{center}
\begin{tabular}{|c|c|c|c|} 
\hline
Parameter            &    Result    \\ \hline \hline
$\Gs~[\invps ]$      &  $0.680   \pm 0.034  \pm 0.027 $ \\
$\DGs~[\invps ] $    &  $0.084   \pm 0.112 \pm 0.021 $ \\
$|A_{\perp}|^{2}$       &  $0.279   \pm 0.057 \pm 0.014 $ \\
$|A_{0}|^{2}$          &  $0.532   \pm 0.040 \pm 0.028$ \\
$\cos\delta_{\parallel}$ &  $-1.24    \pm 0.27 \pm 0.09 $ \\
\hline 
\end{tabular}
\end{center}
\caption{ Parameters extracted from the fit to the \BsJphi\ events in the 
untagged analysis. 
The first uncertainty is statistical and the second is systematic.
LHCb preliminary results from 36 \invpb.}
\label{tab:untagged}
\end{table}

\begin{figure}[htb]
\centering
\includegraphics[width=0.4\textwidth]{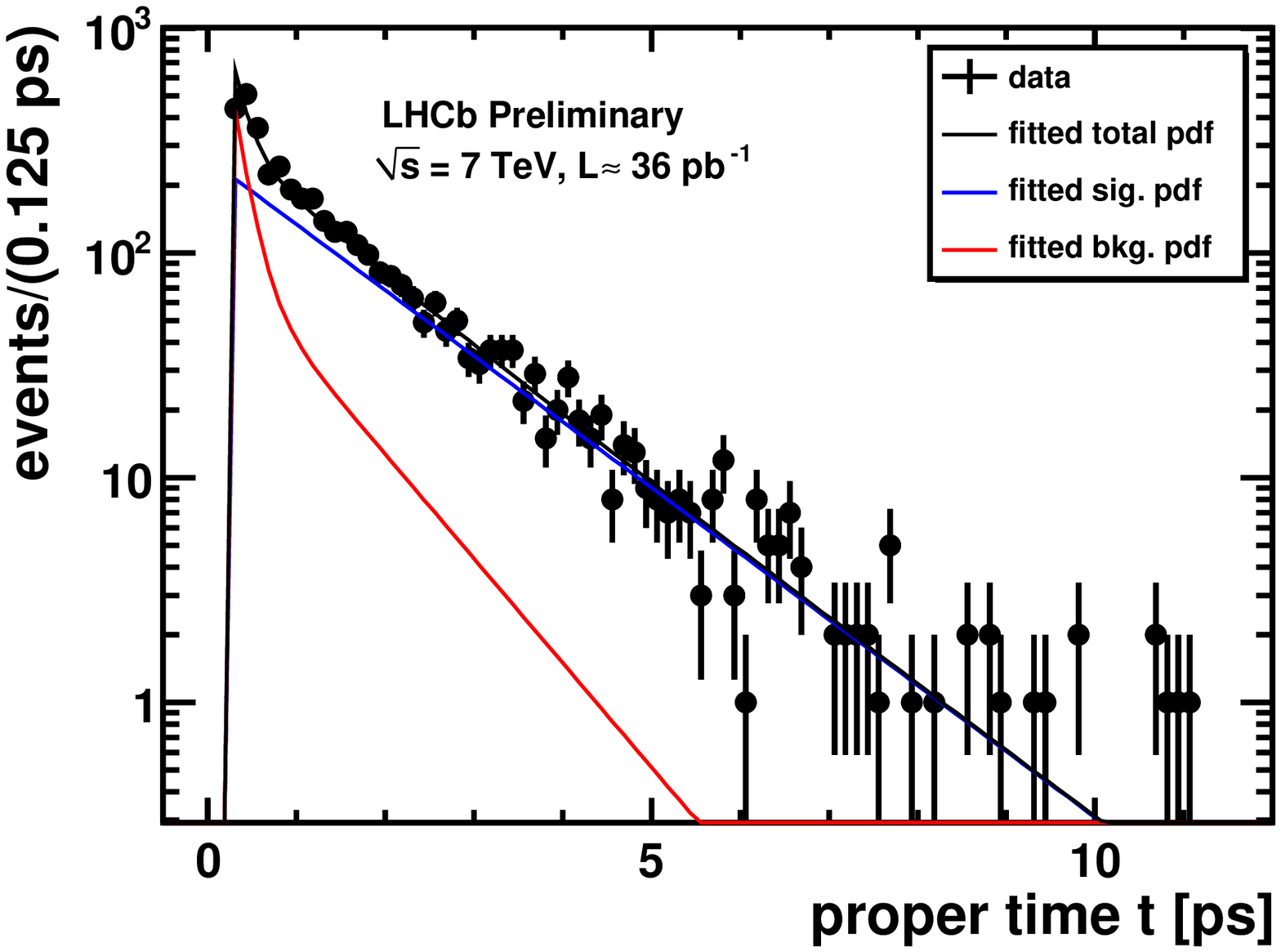}
\includegraphics[width=0.4\textwidth]{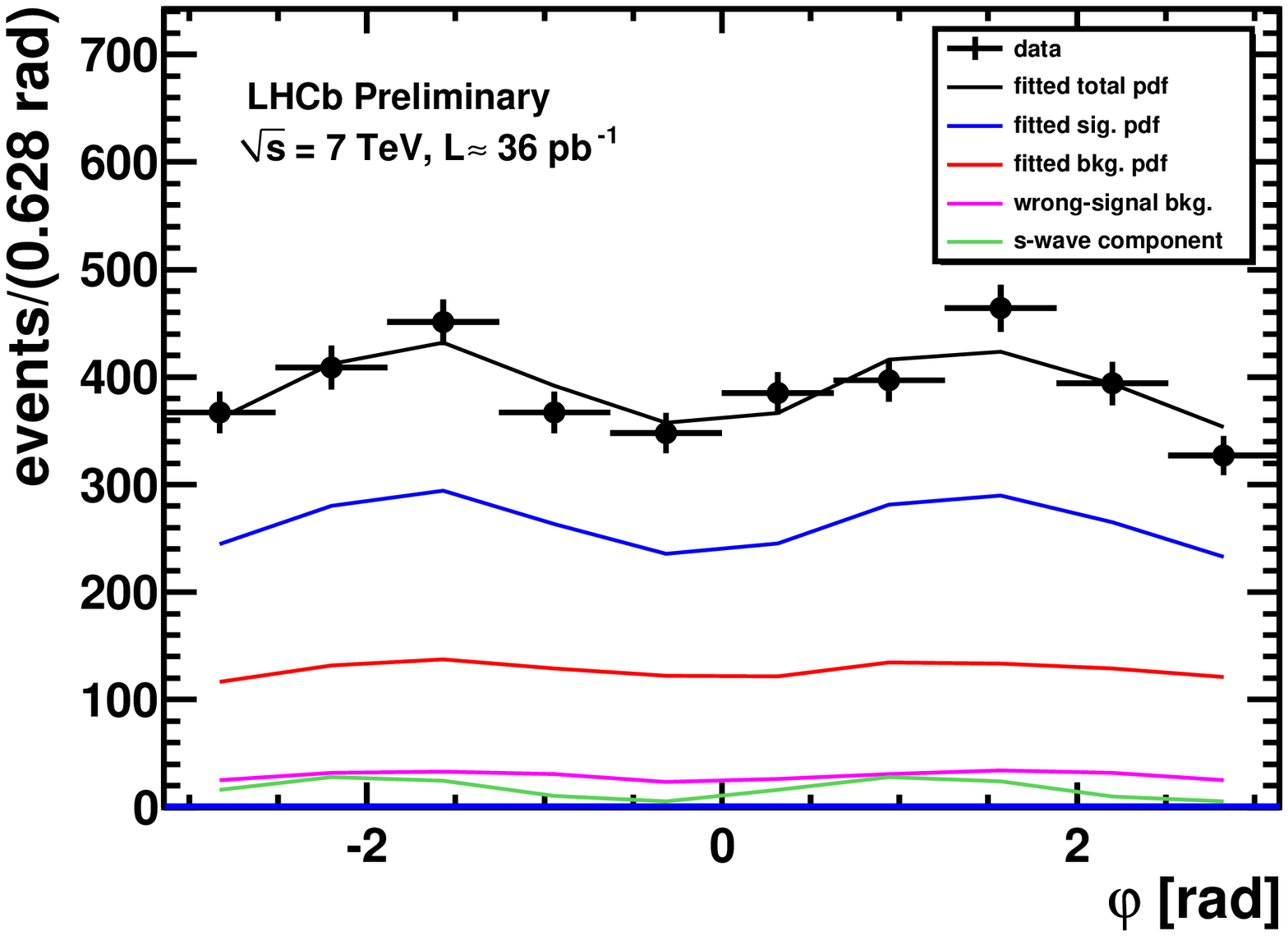}\\
\vskip 0.5cm
\includegraphics[width=0.4\textwidth]{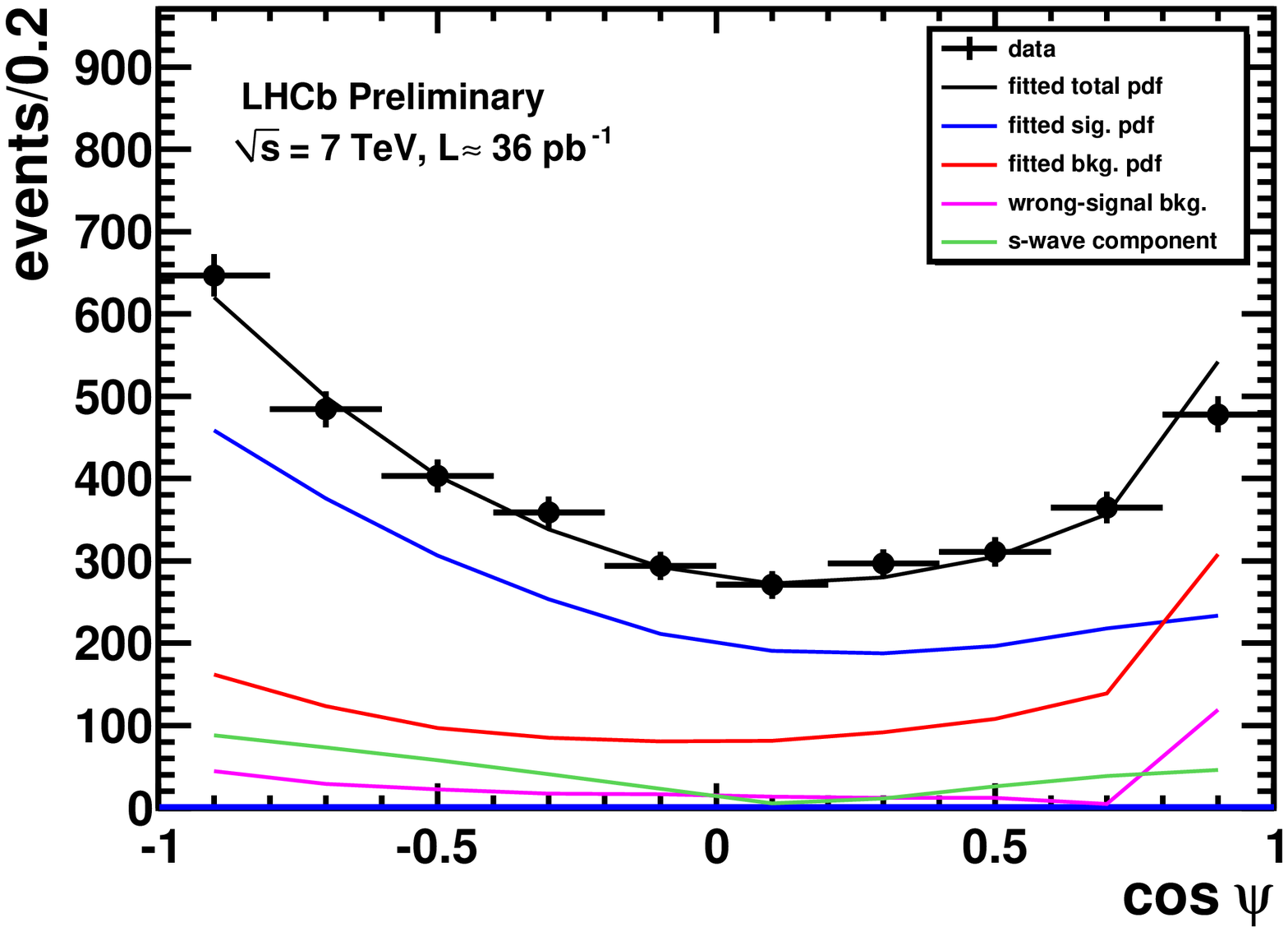}
\includegraphics[width=0.4\textwidth]{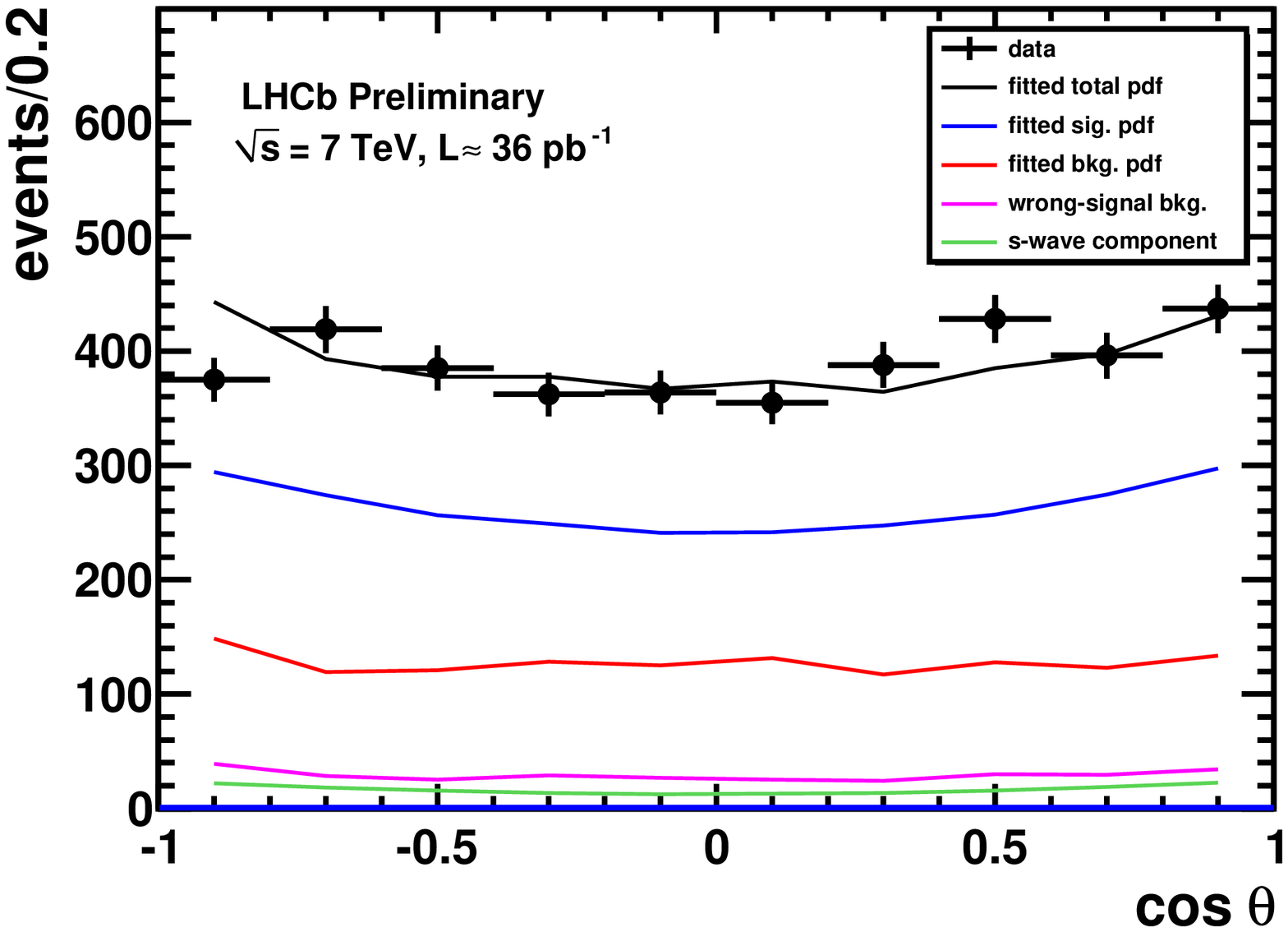}
\caption{Fitted PDF projected on the the proper-time and the transversity angles
compared to the data distributions for the  selected \BdJKst\ candidates.}
\label{fig:JpsiKstar}
\end{figure}

The \BdJKst\  decay channel provides a valuable check for the \BsJphi\ 
angular analysis since it is also a pseudo-scalar to vector-vector decay which 
occurs via similar 
(parity-odd and parity-even) decay amplitudes which are already well 
measured~\cite{ref:otherJK}.
The final state is in this case flavour specific, with the kaon charge 
identifying the flavour of the decaying neutral B meson. 
A full fit to the mass, proper-time and angular distributions is performed
to determine the polarisation amplitudes $\aparO$, $\aperpO$ and the 
related strong phases $\delpar$ and $\delperp$ for the decay.
Results are shown in Table~\ref{tab:JpsiKstar} and 
Fig.~\ref{fig:JpsiKstar}~\cite{ref:note2}.
They are in agreement with previous measurements, even if, with the present event sample,
not yet competitive with them.

\begin{table}[!hbtp]
\begin{center}
\begin{tabular}{|c|c|} \hline
Parameter &Result  \\\hline \hline
$|A_{\parallel}|^{2}$ &$0.252\pm 0.020\pm 0.016$ \\
$|A_{\perp}|^{2}$    &$0.178\pm 0.022\pm 0.017 $\\
$\delta_{\parallel}$ (\rad)&$-2.87\pm 0.11\pm 0.010$\\
$\delta_{\perp}$ (\rad)    &$3.02\pm 0.10\pm 0.007$ \\
\hline
\end{tabular}
\end{center}
\caption{Parameters extracted the fit to the selected \BdJKst\ events.
The first uncertainty is statistical and the second is systematic.
LHCb preliminary results from 36 \invpb.}
\label{tab:JpsiKstar}
\end{table}

In a second \BsJphi\ analysis flavour tagging is used ({\it tagged analysis}). 
In LHCb the flavour of the B meson at production is determined by several algorithms, 
the opposite side (OS) taggers uses four different signatures, namely high $p_T$ 
muons, electrons and kaons and the charge of an inclusively reconstructed 
secondary vertex. The same side (SS) taggers use kaons and pions for \Bs\
or \Bd\ and \Bu\ mesons, respectively.
The algorithms have been optimized on data for the maximum tagging power 
$\epsilon D^2$, where $\epsilon$ is the tagging efficiency, $D=1-2\omega$ 
is the dilution and $\omega$ the mistag. 
\BuJK\ and \BdDstmu\ have been used as control channels for the OS taggers
~\cite{ref:note3}. 
The tagging power is enhanced by using in the fits a per-event mistag probability, 
this is obtained by the combination of the different taggers and is  
calibrated on data, using \BuJK\ events, and validated on  \BdJKst.
The tagging power for \BsJphi\ events is found to be $\epsilon D^2=2.2\pm0.5$ \%,
where the uncertainty is dominated by the statistical fluctuations in the control channel.
To calibrate  the SS kaon tagger \BsDspi\ events are used, but the data sample 
from the 2010 run is too small and so this tagger has not been considered 
for the time being.
  
\begin{figure}[htb]
\centering
\includegraphics[width=0.6\textwidth]{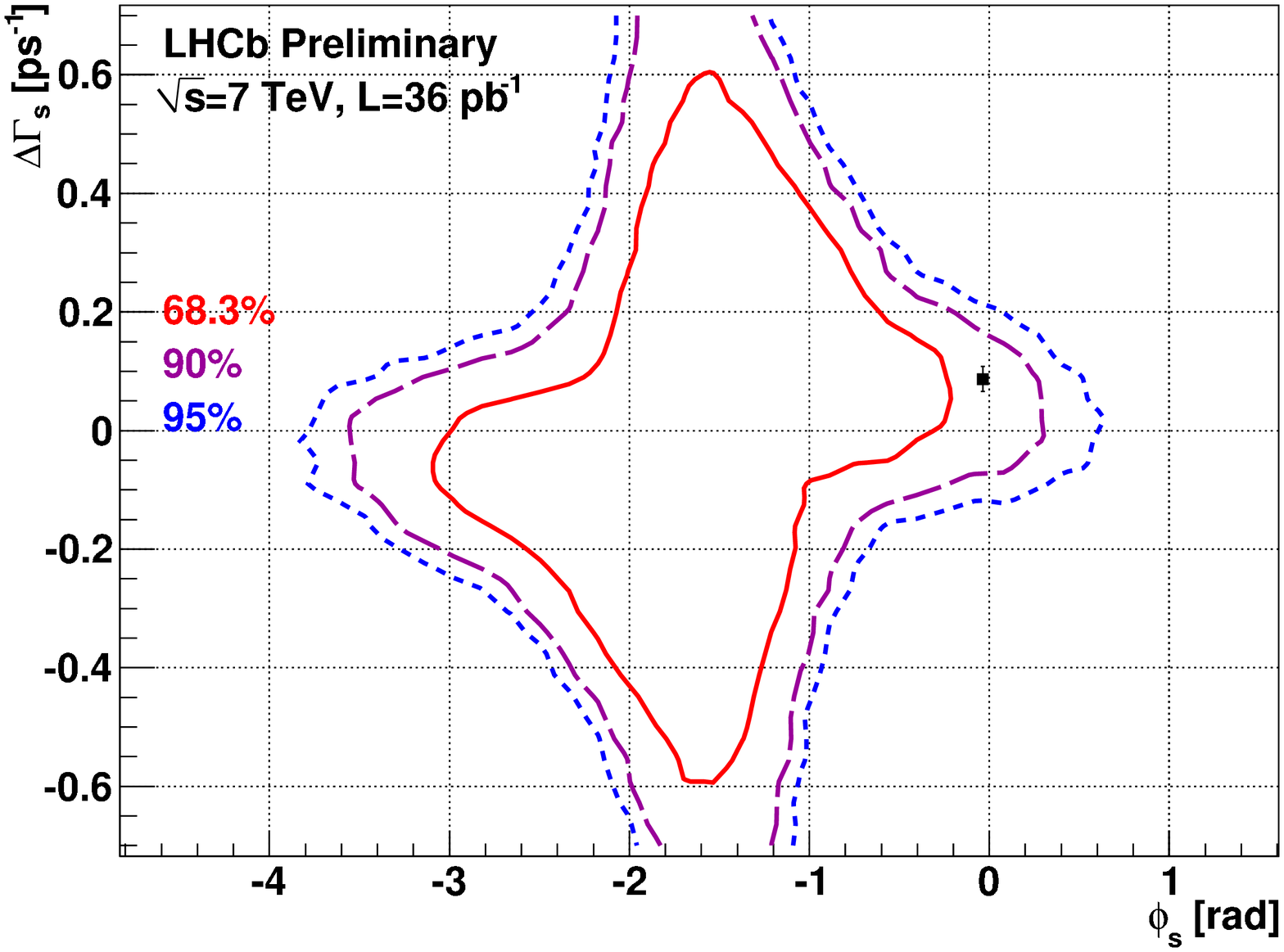}
\caption{ Feldman and Cousins regions in the \DGs\ -\myphis\ plane. The CL at the 
SM point (black square) is 0.785 which corresponds to a deviation of 1.2 $\sigma$.}
\label{fig:contour}
\end{figure}

For the tagged analysis events selected by a displaced track trigger 
(decay-time biased) have been also used, for a total sample of 757$\pm$28 signal
events ($t>0.3$ ps) ~\cite{ref:note6}.
The time acceptance for the decay-time biased events is obtained from data 
using the time distribution of events which pass both trigger conditions. 
The result of the fit is presented as two-dimensional confidence level regions in the 
\DGs\ -\myphis\ plane obtained using a likelihood ratio ordering, 
following the prescription of Feldman and Cousins~\cite{ref:Feldman}. 
Fig.~\ref{fig:contour} shows the 68.3\%, 90\% and 95\% confidence level contours 
in the \DGs\ -\myphis\ plane. 
With current event yields systematic uncertainties were found to 
have only a small effect on the contours. 
Therefore the contours include only the statistical 
uncertainties, with the exception of the uncertainties due to flavour tagging calibration 
parameters and mixing frequency, which were floated in the fit. 
When projecting the confidence level contours onto one dimension it results 
$-2.7< \myphis <0.5$ \rad\ at 68\% CL.
Substantial improvements are expected in the determination of \myphis\ 
with the larger 2011 data sample and the use of SS kaon tagging.

The flavour tagging is used also in the measurement of time-dependent CP violation 
in \BdJKS\ decays~\cite{ref:note4}.
The preliminary result is $\sin 2\beta \simeq S = 0.53^{+0.28}_{-0.29}\pm0.07$, where 
the first uncertainty is statistical and the second is systematic.
The data sample is currently too small for a measurement competitive with B factories, 
but it adds a valid demonstration of LHCb capability in time-dependent CP analysis.
Moreover,  with the integrated luminosity foreseen over the coming few years LHCb
will be able to make a precise determination of this important parameter.

\section{Measurement of the \Bs\ oscillation frequency}

\begin{figure}[htb]
\centering
\includegraphics[width=0.39\textwidth]{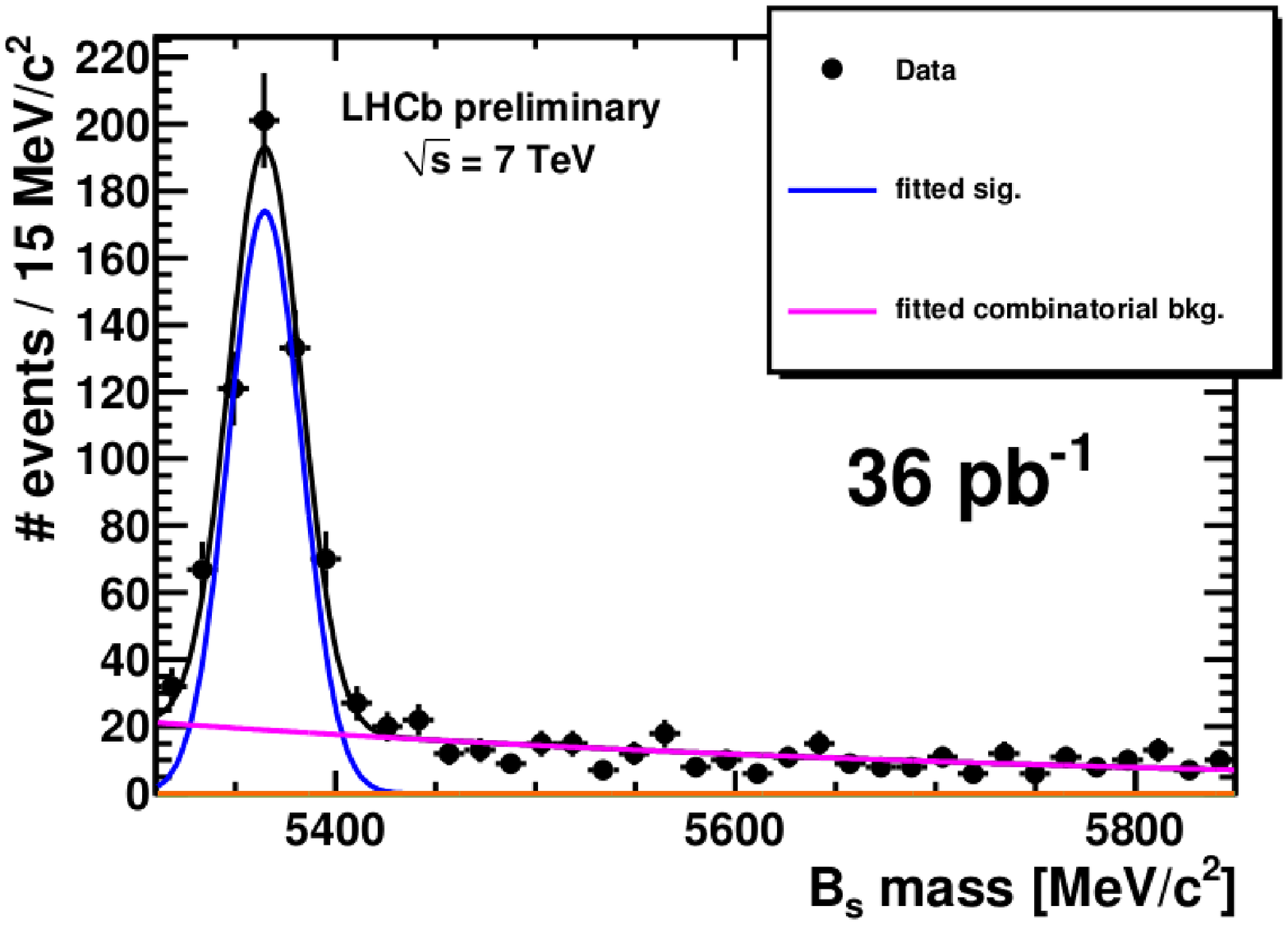}
\includegraphics[width=0.39\textwidth]{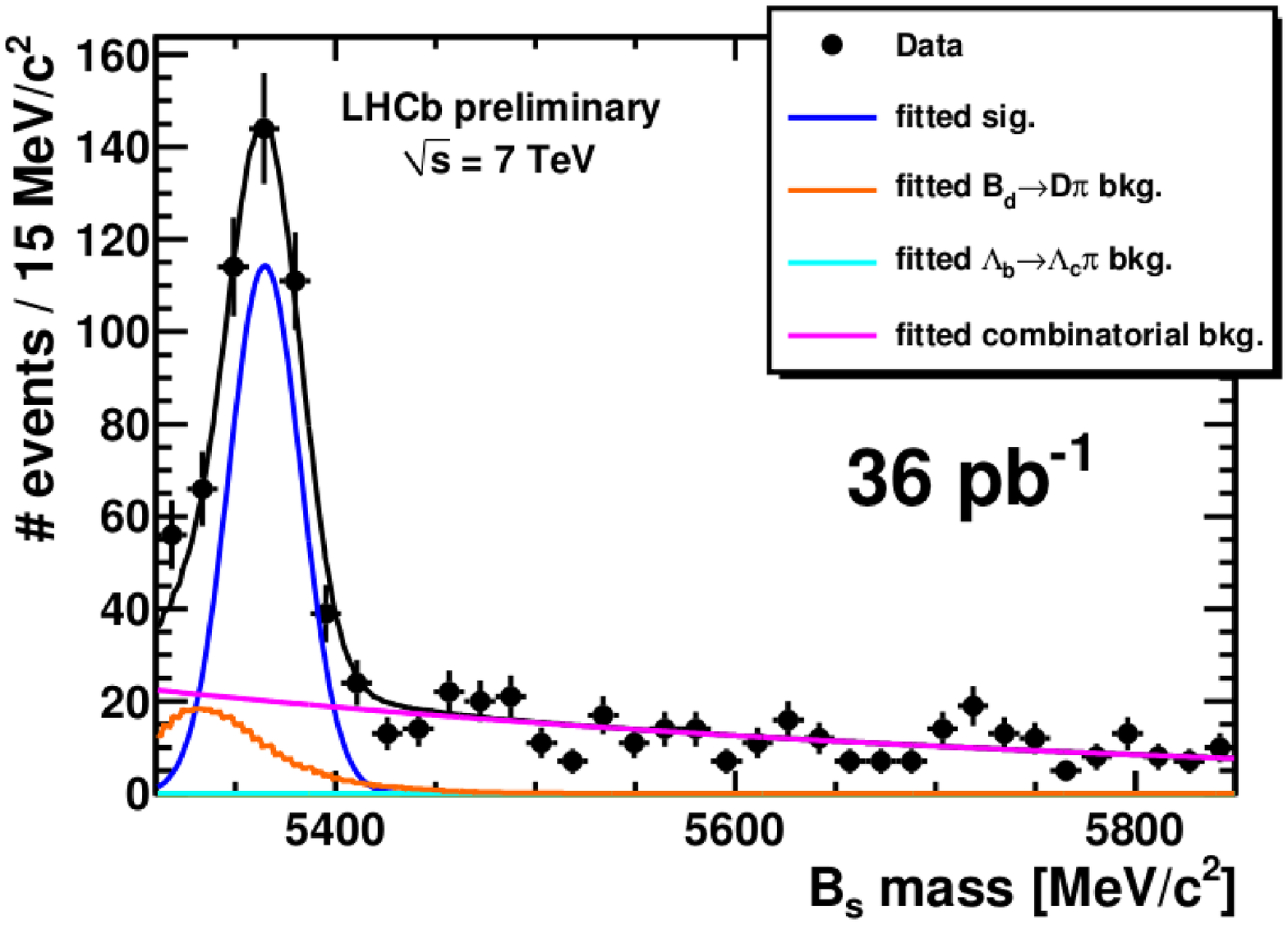}\\
\includegraphics[width=0.39\textwidth]{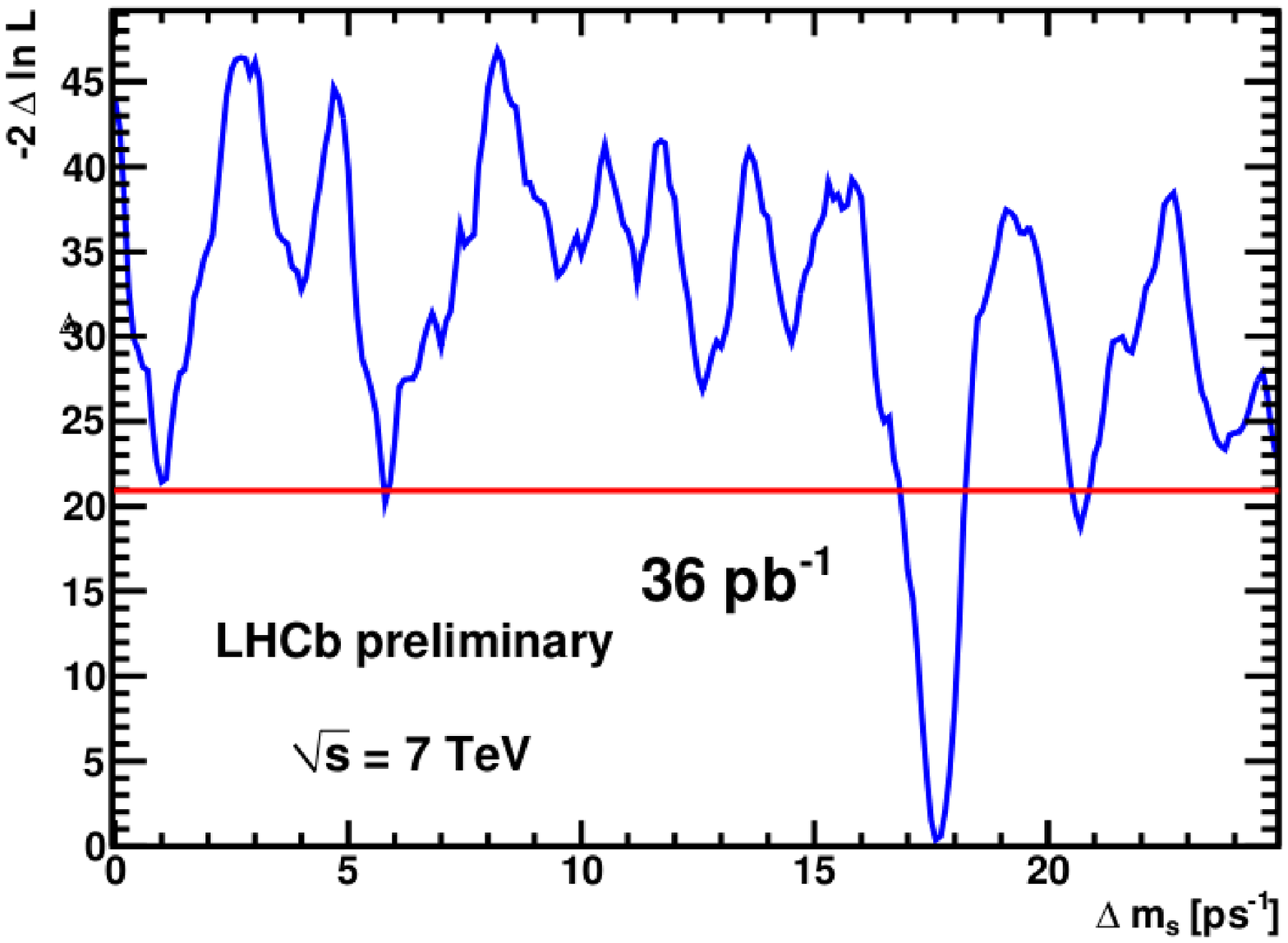}
\includegraphics[width=0.39\textwidth]{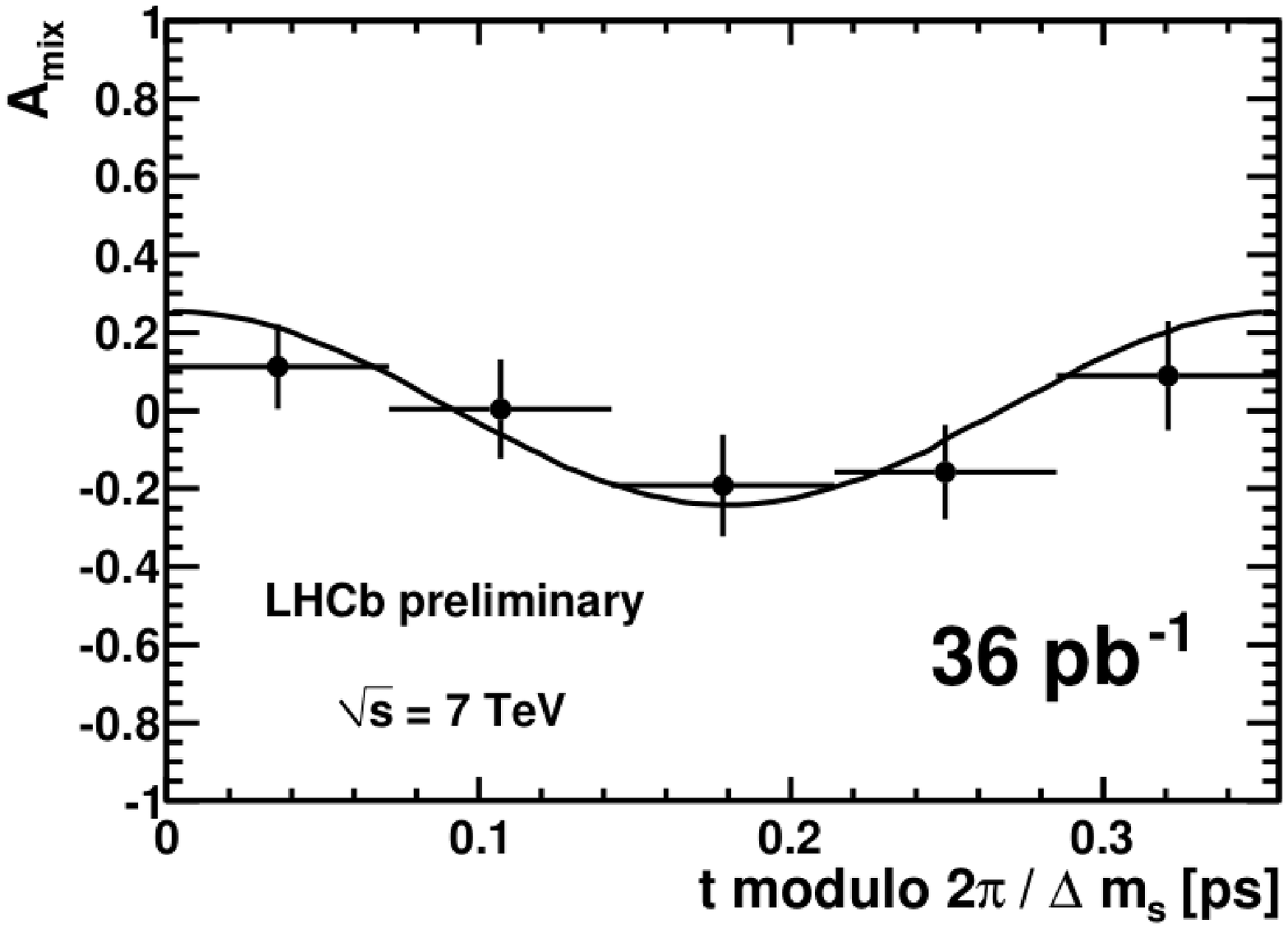}
\caption{Top: fit to the mass distributions of 
$B_s \rightarrow  D^-_s(\phi \pi^-) \pi^+$ (left) and
$B_s \rightarrow D^-_s (K^*K) \pi^+$ (right) events.
Bottom left: likelihood scan for \dms\, the line indicates the likelihood value
evaluated in the limit of infinite mixing frequency.
Bottom right: mixing asymmetry for signal \Bs\ candidates as a function of 
proper-time modulo $2\pi / \dms$.}
\label{fig:mixing}
\end{figure}

For the measurement of the \Bs\ oscillation frequency a  total of 1350 \Bs\ signal 
candidates are reconstructed in four different \Bs\ decay modes, 
as indicated in Table~\ref{tab:Bsmodes}~\cite{ref:note5}. 
The invariant mass distributions of two of these modes are shown in Fig.~\ref{fig:mixing}.
The four modes are analysed individually but a single set of physical parameters, 
namely \Bs\ mass, lifetime and mixing frequency, is determined in a common fit 
to mass and proper-time distributions of all modes. 
The true proper-time is smeared by detector resolution $\sigma_t$, 
and good proper-time resolution is crucial for resolving fast \Bs\ oscillations. 
In this analysis the event-per-event resolution $\sigma_t$ is used, 
it is calibrated on data using \Bs\ candidates 
formed by a prompt \Dsm\ and a pion, obtaining an average proper-time resolutions of 
44 and 36 fs for \BsDspi\ and \BsDstpi\ decays, respectively.
OS flavour taggers are used with a total tagging power of
$\epsilon D^2=3.8\pm2.1$\%.

\begin{table}[!hbtp]
\begin{center}
\begin{tabular}{|c|c|} \hline
decay mode                           & \# signal candidates \\ \hline
$B_s \rightarrow  D^-_s(\phi \pi^-) \pi^+$     & 515 $\pm$ 25 \\
$B_s \rightarrow D^-_s (K^*K) \pi^+$           & 338 $\pm$ 27 \\
$B_s \rightarrow D^-_s(K^+K^-\pi^-) \pi^+$     & 283 $\pm$ 27 \\
$B_s \rightarrow D^-_s (K^+K^-\pi^-)3\pi$      & 245 $\pm$ 46 \\ \hline
\end{tabular}
\caption{Number of $B^0_s$ signal candidates. }
\label{tab:Bsmodes}
\end{center}
\end{table}

The mixing frequency is measured as \dms\ = $17.63 \pm 0.11 \pm 0.04$ ps $^{-1}$
with a 4.2 $\sigma$ significance.
Fig.~\ref{fig:mixing} shows the likelihood profile as a function of \dms\ and the 
mixing asymmetry for signal \Bs\ candidates as a function of proper-time.

\section{Measurement of \BsKK lifetime}

Two-body charmless B decays offer a rich phenomenology to explore the
phase structure of the CKM matrix and to search for manifestations of
New Physics beyond the SM.  The lifetime of the decay \BsKK is of considerable 
theoretical interest to discover NP. 
This decay channel is particularly suitable to measure with the LHCb
experiment due to its powerful pion, kaon and proton
identification capabilities.
The proper time distribution of the untagged \BsKK decay is given by
	\begin{equation}
	\hat{\Gamma}(\BsKK) \equiv R_H e^{-\Gamma_H t} + R_L e^{-\Gamma_L t},
	\end{equation}
where $R_H$ and $R_L$ are the fractions of the heavy ($\Gamma_H$) and
light ($\Gamma_L$) states contributing to the \BsKK
decay. $\hat{\Gamma}(\BsKK)$ is primarily sensitive to the width of
the short-lived light state of the \Bs.  

Comparing the measurement with a lifetime obtained from a flavour
specific decay allows \DGs\ and \DGs/2\Gs\ to be extracted
inside the SM. Any NP effects in \BsKK\ will lead to a
reduction in the measured \DGs\ from the SM value~\cite{ref:Grossman}.

Two measurements of the lifetime have been performed with
first LHCb data~\cite{ref:note18}.
The selection procedure of the sample of B mesons decaying into two hadrons
makes minimum requirements on the flight distance of the B meson,
as a consequence it tends to reject candidates which decay after a short proper time. 
Two independent data-driven approaches have been developed to compensate
for the resulting bias.  One extracts the acceptance function from the
data, and then applies this acceptance correction to obtain a
measurement of the \BsKK\ lifetime.  The other method cancels the
selection bias by taking a ratio of the \BsKK\ and \BdKpi\
proper decay time distributions, exploiting the fact that 
the two channels have very similar decay topologies.
The \BsKK lifetime is then extracted,
using the world average measurement of the \BdKpi lifetime as input.
The results of the two measurements are in good agreement and are combined.  
The lifetime is measured as 
$\tau_{B_s} = 1.440 \pm 0.096\pm0.010  \ps$
where the first uncertainty is statistical and the second is systematic.
It represents the world best precision on this quantity.	

\section{Conclusion}
Excellent performance of LHCb in time-dependent measurements have been 
proved on 2010 data. All the steps towards a \myphis\ measurement with
\BsJphi\ decays have been completed and the world best determination
of the mixing phase is expected with 2011 LHCb data.
The preliminary measurement of the mixing frequency \dms\ in \BsDspi\  decays 
has been obtained with a precision competitive with existing results.
Many other decay modes of B mesons are under study which will provide
precise measurements of CP violation and constraints on NP models.

\end{document}